\documentclass[fleqn,usenatbib]{mnras}
\usepackage[T1]{fontenc}
\usepackage{natbib, amsmath, graphicx,newtxmath}
\usepackage{pifont}

\title[Baryon Painting with Image-to-Image Neural Networks]{Painting baryons onto $N$-body simulations of galaxy clusters with image-to-image deep learning}

\author[Chadayammuri et al]{
Urmila Chadayammuri$^{1}$\thanks{E-mail: uchadayammuri@cfa.harvard.edu}, Michelle Ntampaka$^{2, 3}$, John ZuHone$^{1}$, \'Akos Bogd\'an$^{1}$, 
\newauthor  Ralph P. Kraft $^{1}$
\\
$^{1}$Center for Astrophysics | Harvard \& Smithsonian, 60 Garden Street, Cambridge, MA 02140, USA\\
$^{2}$Space Telescope Science Institute, 3700 San Martin Drive, Baltimore, MD 21218, USA\\
$^{3}$Department of Physics and Astronomy, Johns Hopkins University, 3400 North Charles Street, Baltimore, MD 21218, USA
}

\pubyear{2023}

\begin{document}
\label{firstpage}
\pagerange{\pageref{firstpage}--\pageref{lastpage}}
\maketitle

\begin{abstract}
Galaxy cluster mass functions are a function of cosmology, but mass is not a direct observable, and systematic errors abound in all its observable proxies. Mass-free inference can bypass this challenge, but it requires large suites of simulations spanning a range of cosmologies and models for directly observable quantities. In this work, we devise a U-net \textemdash{} an image-to-image machine learning algorithm \textemdash{}  to ``paint'' the IllustrisTNG model of baryons onto dark-matter-only simulations of galaxy clusters. Using 761 galaxy clusters with $M_{200c} \gtrsim 10^{14}M_\odot$ from the TNG300 simulation at $z<1$, we train the algorithm to read in maps of projected dark matter mass and output maps of projected gas density, temperature, and X-ray flux. Despite being trained on individual images, the model reproduces the true scaling relation and scatter for the $M_{DM}-L_X$, as well as the distribution functions of the cluster X-ray luminosity and gas mass. For just one decade in cluster mass, the model reproduces three orders of magnitude in $L_X$. The model is biased slightly high when using dark matter maps from the DMO simulation. The model performs well on inputs from TNG300-2, whose mass resolution is 8 times coarser; further degrading the resolution biases the predicted luminosity function high. We conclude that U-net-based baryon painting is a promising technique to build large simulated cluster catalogs which can be used to improve cluster cosmology by combining existing full-physics and large $N$-body simulations. 
\end{abstract}

\begin{keywords} galaxies: clusters: intracluster medium --  cosmology: large-scale structure of Universe -- machine learning
\end{keywords}

\section{Introduction} 
\label{sec:main}

In any cosmological model, the growth of cosmological structure is driven by the interplay between the gravitational collapse of the dark and baryonic matter on the one hand and Hubble expansion and acceleration by dark energy on the other \citep{Frenk1988}. In particular, the mass function of galaxy clusters, the largest virialised structures in the Universe today, is a concrete prediction of a given cosmological model as a function of cosmic time \citep[][and references therein]{MoWhite1996, Bryan1998,Voit2005, Tinker2008}. The latest generation of $N$-body simulations predicts the cluster mass function for hundreds, or up to tens of thousands, of different values of key cosmological parameters - the relative energy densities of matter $\Omega_m$ and dark energy $\Omega_\Lambda$, the normalisation of the matter power spectrum at scales of 8 Mpc/$h$ $\sigma_8$, the dark energy equation of state $w$, and the spectral index of the primordial overdensity spectrum $n_s$ - by simulating only gravitational interactions over extremely large volumes \citep[e.g.][]{Prada2012,Bhattacharya2013, Klypin2016, Villaescusa2020,Maksimova2021, Ishiyama2021}. This process has been accelerated even further by the advent of differentiable simulations \citep{Modi2021, Li2022}, and is crucial for enabling likelihood-free or simulation-based inference \citep[see, e.g.,][and references therein]{Alsing2019,Cranmer2020}.

Since they probe the growth of structure, galaxy clusters provide cosmological constraints almost orthogonal to geometric measurements such as the Cosmic Microwave Background \citep[CMB, ][]{Planck2014, Planck2020}, Type Ia supernovae \citep[SNIa][]{Astier2006} and Baryon Acoustic Oscillations \citep[BAO][]{Alam2017}. Improving cluster cosmology constraints thus dramatically improves the error bars on cosmological parameters, especially on the clustering parameter $S_8 = \sigma_8\left(\Omega_m/0.3\right)^{1/4}$, when combined with the geometric methods \citep{Pillepich2018b-eROSITA}. However, crucial systematic issues remain in deriving cosmological parameters from galaxy cluster observations.

One issue is that the cluster mass function is systematically affected by baryonic effects, which are not accounted for in analytic models or $N$-body simulations. Radiative cooling, star formation, feedback from stars and supermassive black holes, cosmic rays, and magnetic fields all act to modify the cluster population. Incorporating all these processes requires very computationally expensive and high resolution (magneto-)hydrodynamical simulations of cosmological volumes; only a few dozen of these exist today \citep{Borgani2004, Nagai2007, LeBrun2014, Planelles2014, Schaye2015,Dolag2016,Suto2017-HorizonAGN, Tremmel2017, Dave2019, Pillepich2018a-TNG}, most using either \citet{Planck2016} or WMAP-7 \citet{Komatsu2011} cosmology. These have shown that simply including baryons changes the measured $\Omega_m$ and $S_8$ from an X-ray selected cluster sample by 4$\%$ and 12$\%$, respectively \citep{Bocquet2016, Castro2021}, even if the cluster masses are measured perfectly. The result holds even for lensing surveys, highlighting that the baryons affect the dark matter halos themselves \citep{Ferlito2023}.

Second, measuring cluster masses is far from trivial. In galaxy clusters, for example, the space between galaxies is filled with a plasma, known as the intracluster medium (ICM). To first order, this plasma is in hydrostatic equilibrium with the total gravitational potential, so that it can be used to infer the total mass; subtracting the directly observed gas mass then yields the dark matter mass. Most commonly, this is done via the X-ray emission from the cooling cluster electrons \citep{Reiprich2002, Vikhlinin2006} and inverse compton (IC) scattering of CMB photons, known as the Sunyaev-Zel'dovich effect \citep[SZE][]{1972CoASP...4..173S}.
However, the ICM also experiences other processes, like radiative cooling, subcluster- and cluster-cluster mergers, and jet and radiative feedback from stellar evolution \citep{Battaglia2012} and supermassive black holes (SMBH) \citep[][and references therein]{McNamara2007, Battaglia2010, McCarthy2011}. In the central regions, the ICM deviates from gravity-only (a.k.a. self-similar) predictions primarily due to active galactic nucleus (AGN) feedback, while in the outskirts it is due to clumpiness and cosmological accretion. In summary, the intracluster medium is an imperfect tracer of the dark matter potential. 

For a few dozen low- and intermediate-redshift, high-mass, dynamically relaxed clusters, it is possible to measure the total mass from X-ray emission profiles under the assumption of hydrostatic equilibrium \citep{Vikhlinin2006}. These are used to construct scaling relations, which are then extrapolated to less massive and/or more distant clusters, potentially viewed with lower resolution. Even studies that use only visually relaxed clusters, which are expected to be in hydrostatic equilibrium \citep{Allen2002,Ettori2019}, are known to suffer from hydrostatic bias at the 5-10$\%$ level \citep{Meneghetti2010}. The mass scaling relation of even relatively good mass proxies like the X-ray Compton-like parameter $Y_X = M_{gas}\times T_X$ \citep{Kravtsov2006} has an intrinsic scatter of 0.3~dex \citep{Chiu2022}, which adds 40$\%$ uncertainty to cosmological parameter estimates \citep{Planck2014}. The masses are often calibrated using gravitational lensing \citep{Allen2001, Hoekstra2012, Mahdavi2013,vdLinden2014, Merten2015, Mantz2015}, although it has been shown that weak lensing carries an intrinsic scatter of $20\%$ at cluster mass scales due in part to halo triaxility, before even incorporating baryonic effects \citep{Becker2011}. The net result of the scatter and biases is that is that the cluster mass conversion varies widely in the X-ray literature and produces cosmological parameter estimates that differ from each other by up to 2.5$\sigma$. As a result, studies like \citet{Planck2020} have avoided using clusters for cosmological constraints altogether. 

The dominant approach to improving cluster cosmology involves reducing the scatter and/or bias in the scaling relations between X-ray or SZE observables and cluster mass \citep{Shi2016}. Cutting out the central 0.1-0.15$R_{500}$ of the ICM, for example, has proved an effective way to do so \citep{Maughan2007}. However, this is only possible for nearby clusters observed with high-resolution instruments like \textit{Chandra} and \textit{XMM-Newton}, whereas survey telescopes like \textit{eROSITA} have a much lower resolution, making it unfeasible to reliably mask out only photons associated with cluster cores. Since \textit{eROSITA} is expected to find $\sim10^5$ clusters with virial mass $M_{vir} = M_{200c} > 10^{13}M_\odot$, the bulk of them at lower masses, higher redshifts and/or with shallow exposures \citep{Predehl2021}, it is crucial to use all the information from the X-ray images to test cosmological models. The same can be said of upcoming SZE surveys like CMB-S4 and the Simons Observatory, which will have beam sizes of order $1'$ \citep{Abazajian2016,Ade2019,Abazajian2019}. Restricting analyses to low-redshift, relaxed objects would eliminate the bulk of the unprecedented cluster sample obtained by these long-anticipated surveys.  

A crucial, complementary approach is improving the incorporation of baryons into theoretical predictions. In the field of small-scale, near-field cosmology, a long-standing ``problem'' of missing satellites, numerous small-scale halos predicted by dark-matter-only simulations of Milky-Way like systems, can be solved entirely by incorporating baryonic feedback into the simulations \citep{Brooks2013, DelPopolo2014}, although more exotic solutions continue to be proposed. While baryonic feedback cannot disrupt cluster-scale halos in the same way, it can certainly affect their appearance at X-ray and SZE wavelengths by significantly reshaping the diffuse intracluster medium \citep{Nagai2007, Martizzi2012, Bryan2013, Bocquet2016, Castro2021}. Two bottlenecks stand in the way of incorporating baryons into cosmological simulations. First is the computational infeasibility of running many large-volume boxes, each with a different cosmology and large enough to contain a significant number of galaxy clusters while simultaneously implementing baryonic physics. Second is the uncertainty in the baryonic models themselves - no simulation so far has reproduced every observation of galaxy properties and their evolution over time.

``Baryon painting'' is the post-processing of dark-matter-only simulations to capture the net effect of the baryons, had they been implemented directly. Since it uses properties already computed by the N-body simulations, baryon painting is an extremely cheap process. This addresses both the bottlenecks above - it allows us to paint baryons following many different prescriptions, and onto many N-body simulations, for a marginal computational cost.

Baryon painting has already been undertaken using (semi-)analytic prescriptions that map halo properties to baryonic observables in hydrodynamic simulations \citep{Lu2022, Osato2023,Williams2023, Keruzore2023, Zhong2023}. However, such halo-based models assume that the baryons in a halo have always been dynamically related to the dark matter in the same halo, whereas studies have shown that they in fact carry dynamical information from much further out in the cosmic web, from where they were transported in \citep{Kimm2011, Liao2017}. 

\begin{figure*}
    \includegraphics[width=\textwidth]{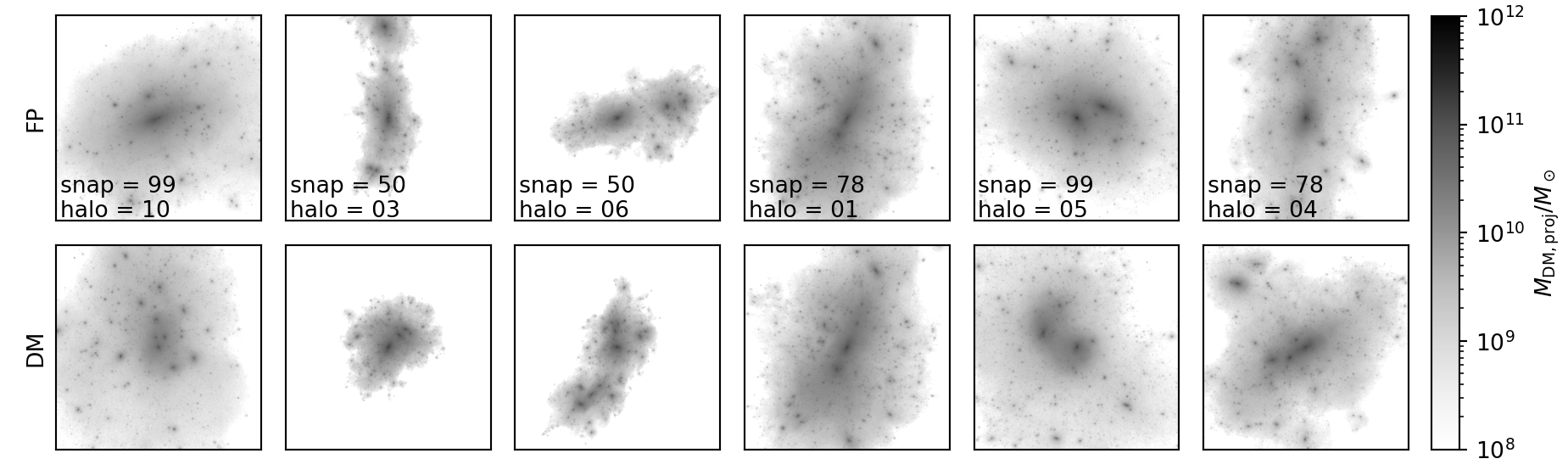}
    \caption{Examples of dark matter halos from the full-physics (FP) run and their counterparts in the dark-matter-only (DMO) run. The color shows the projected dark matter mass along the line of sight. The halos have been matched bijectively (i.e. in both directions, DMO $\leftrightarrow$ FP) by tracing particle IDs from the initial conditions, i.e. these halos contain mostly the same DM particles from the initial snapshot. Nevertheless, they look very different by z$\lesssim$1, due to the chaotic nature of the N-body problem.}
    \label{fig:dm_vd_fp}
\end{figure*}
Machine learning offers a powerful new toolkit to target the problem of painting baryons onto dark-matter-only simulations. Convolutional neural networks (CNNs) are particularly good at extracting complex features from multi-dimensional inputs by learning a series of filters in order to minimise the error in predicting known properties of the training sample. They have been used extensively to predict cluster masses from mock observations \citep{Ntampaka2015, Ntampaka2016,Ho2019,Ntampaka2019, Gupta2020}. Generative neural networks have been used to expand samples of galaxy cluster SZE maps given a sample from one simulation \citep{Troster2019}; they can also predict SZE maps using halo properties from dark-matter-only simulations \citep{Thiele2020,Thiele2022}. \cite{deAndres2023} trained a variety of random forest and gradient boost algorithms to predict baryonic properties from The Three Hundred simulations \citep{Cui2018} using a data vector of 26 quantities from the corresponding halo in the MDPL $N$-body simulation \citep{Klypin2016} as input. They were able to recover gas mass, mass-weighted gas temperature and other baryonic properties with root-mean-squared errors of 4-8$\%$. Using a similar technique of using vectors of halo properties to predict observable signals, \citet{Delgado2023} and \citet{Pandey2023} quantified how varying baryonic physics affects the matter power spectrum in the CAMELS simulations \citep{Villaescusa2022}.

We would like to reproduce not only the mean mapping, but also the scatter and the diversity of the cluster population, so as to produce distribution functions of and scaling relations between directly observable quantities from dark-matter-only simulations. This will allow direct comparison between the numerous existing $N$-body simulations that explore a wide variety of cosmologies, and observations. \citet{Andrianomena2022} have trained generative adversarial networks (GANs) with 2D images from CAMELS to generate images of gas mass, neutral hydrogen (HI), and magnetic field strength that statistically matched the properties of the training set and encoded the same cosmological information; \citet{Bernardini2022} achieved similar success with images from the FIRE simulations. \citet{Wu2023} was able to predict the stellar masses of galaxies using 2D maps of the dark matter mass. These studies tell us that additional cosmological information is encoded in the spatial distribution of the baryonic and dark matter properties, over and above what can be learned from azimuthally averaged quantities or summary scalars.

In this paper, we aim to combine these two insights - that the baryonic properties can be predicted from dark matter properties, and that cosmological information is encoded in the spatial maps of the dark matter - to train a machine-learning algorithm with 2D images of cluster-scale halos extracted from the magnetohydrodynamic TNG300 simulation. We apply the trained model to dark matter maps from the dark-matter-only simulation run from identical initial conditions to quantify the effect of baryons on the cluster X-ray luminosity and gas mass functions. We quantify the effect of resolution by predicting these distribution functions from lower-resolution runs of the FP simulation. In this way, we present a resolution-calibrated model to paint baryons onto existing $N$-body simulations and set the stage for cluster cosmology from direct observables. We describe the simulations and the projected images in \S \ref{sec:sims}. An overview of CNN-based autencoders and our implementation of one is provided in \S \ref{sec:ml}. We show our results in \S \ref{sec:results}, share caveats and future directions in \S \ref{sec:disc}, and close with conclusions.

\section{Methods}
\subsection{Input simulations}
\label{sec:sims}
The training data for our model comes from the TNG300 simulation \citep{Pillepich2018a-TNG,Nelson2019}. TNG300 is well-suited to our problem in several ways. First, it offers dark-matter-only (DMO) as well as full-physics (FP) runs simulated from identical initial conditions, allowing us to quantify directly the effect of including baryons. Second, its relatively large volume of (205 Mpc/h)$^3$, i.e. $\sim$(302.6 Mpc)$^3$, produces a significant number of galaxy clusters - almost 1000 halos with $M_{200c} > 10^{14}M_\odot$ at $z \lesssim 1$. Projecting these along several viewing angles further amplifies our training sample size. Lastly, the simulation is run from identical initial conditions at 3 different mass resolutions, with the coarser runs matching the resolution of existing large-volume $N$-body simulations. We assess how a model trained on a high-resolution simulation performs on its low-resolution counterpart, i.e. if this is a viable method of super-resolution painting.

The TNG suite uses the moving-mesh code AREPO \citep{Springel2010} to evolve dark-matter particles and gas cells in a cosmological context. Gas can cool through atomic, molecular, metal line and Bremsstrahlung channels; when it meets certain density, temperature and metallicity criteria, it forms star particles, each representing a single stellar population \citep{Pillepich2018a-TNG}. Feedback from supernovae and massive stars is treated in a sub-grid manner. Black holes of mass $8\times10^5M_\odot$ are seeded in Friend-of-Friends (FoF) halos of mass $5\times10^{10}M_\odot$, after which they accrete matter following a modified Bondi-Hoyle prescription; a fraction of the accreted material is reprocessed as kinetic or thermal feedback, depending on the Eddington ratio \citep{Weinberger2017}. TNG is also currently the only high resolution, cosmological volume, hydrodynamical suite to include magnetohydrodynamics. It is certainly the highest resolution cosmological volume simulation suite to include such a wide array of baryonic processes, with softening lengths as low as 250 pc and effective gas cell sizes as small as 47 pc \citep{Nelson2019}. 

Due to the chaotic nature of the $N$-body problem, even with identical initial conditions, the ``same'' dark matter halo looks very different between the DMO and FP runs. Matching halos between the runs is non-trivial, since the halo catalogs are rank ordered by the mass from the Friends-of-Friends catalog at the specific snapshot; this rank ordering can vary subtly due to baryonic effects as well as due to the inherently chaotic nature of the $N$-body problem. Halos between the DMO and FP runs have previously been carefully matched using SubFind \citep{RodriguezGomez2015} and LHaloTree \citep{Nelson2015}, with the latter ensuring a bijective match. Fig \ref{fig:dm_vd_fp} shows pairs of matched halos between the two runs. Despite having similar masses and positions, their spatial structures are often not similar at all, largely due to the chaotic nature of the $N$-body problem. This means that it is not useful to train an algorithm to learn a mapping between baryonic images from the FP run and dark matter properties from the DMO run; these things are not spatially correlated at late times at all. Instead, we break the problem up into two steps - first training between baryons and dark matter in the FP simulation, and then quantifying the systematic differences between the halos in the FP and DMO simulations. 

To create our training sample, we produce maps of the dark matter mass and gas density as an unweighted projection along each of the $x$, $y$ and $z$ axes; thus, the dark matter map shows the projected mass along the line of sight, which can be linearly rescaled to a lensing convergence $\kappa$, and the gas maps are a column density. Temperature is projected with the spectral-like weighting of \citet{Mazzotta2004}, $w_i = n^2T^{-3/4}$, where $n$ is the number density of a gas cell and $T$ its temperature. 

Lastly, we use the \texttt{yt} and \texttt{PyXSIM} packages \citep{Turk2011, ZuHone2016} to generate mock X-ray surface brightness maps of the clusters. The X-ray emission from the ICM is modelled as a thermal plasma described by the Astrophysical Plasma Emission Code (APEC) model \citep{Smith2001}, which depends on the density, temperature, and  metallicity of each hot gas particle $i$, defined as meeting the following criteria:
\begin{align}
    T_i &> 3\times10^5K \\
    SFR_i &= 0 \\
    \rho_g &< 5\times 10^{-25} {\rm g/cm^3}
\end{align}
The TNG suite traces independently the evolution of nine key elements, which produce nearly all of the X-ray emissivity at ICM temperatures and densities \citep{Pillepich2018a-TNG}. The emissivity is predicted for the 0.5-2.0 keV energy range, commonly used in X-ray studies of the ICM. The X-ray emission computed here does not include the AGN luminosity itself, which is expected to be proportional to the instantaneous accretion rate and follow a power-law spectrum rather than APEC \citep{Biffi2018}. 

In principle, using only halo cutouts could produce a bias compared to observational surveys, since there could be contributing emission from other structures along the line of sight. \citet{ZuHone2022} quantified this effect by creating mock X-ray images from halo cutouts and comparing them to full-box projections at a fixed snapshots as well as using a complete light cone, interpolating between simulation snapshots. They showed that using projections from halo cutouts was biased low by only up to $5\%$ compared to projecting the full lightcone, with the bulk of the sample showing much lower bias. Therefore, we opt to ignore the effects of line-of-sight structure and use the simpler model of including only the gas particles that are bound to a given halo. 

The goal is to create high-resolution maps of the gas properties, which can then be passed through a mock observation pipeline to reproduce the resolution of the instrument of choice. We therefore produce simulated cluster images that are 4 Mpc wide and contain 512$x$512 pixels, centered on the gravitational potential minimum of the Friend-of-Friends halo associated with each cutout; the images thus have a resolution of 7.8 kpc/pix. At the 0.5'' resolution of \textit{Chandra}, this corresponds to  2.0/0.6/0.4 pixels at z = 0.1/0.5/1. For a survey telescope like \textit{eROSITA} with an average PSF FWHM of 25', this is far smaller than a single pixel in our images. 

From the TNG300 snapshots at $z = [1.0, 0.5, 0.3, 0.0]$ we extract 761 halos that meet our virial mass criterion of $M_{200c} > 10^{14} M_\odot$. We project each cluster along the $x$, $y$ and $z$ axes, since most clusters are significantly triaxial; thus, we have 2283 sets of images. In principle, CNNs are not invariant to rotation or reflection \citep[e.g.][and references therein]{Zeiler2013}, but we do not add rotations or flips of the data as augmentations since we consider the training sample large enough, and because in practice the clusters are already randomly oriented with respect to the Cartesian axes of the simulation box, so that the lack of rotation does not systematically bias our training sample.

\begin{figure*}
    \centering
    \includegraphics[width=0.8\textwidth]{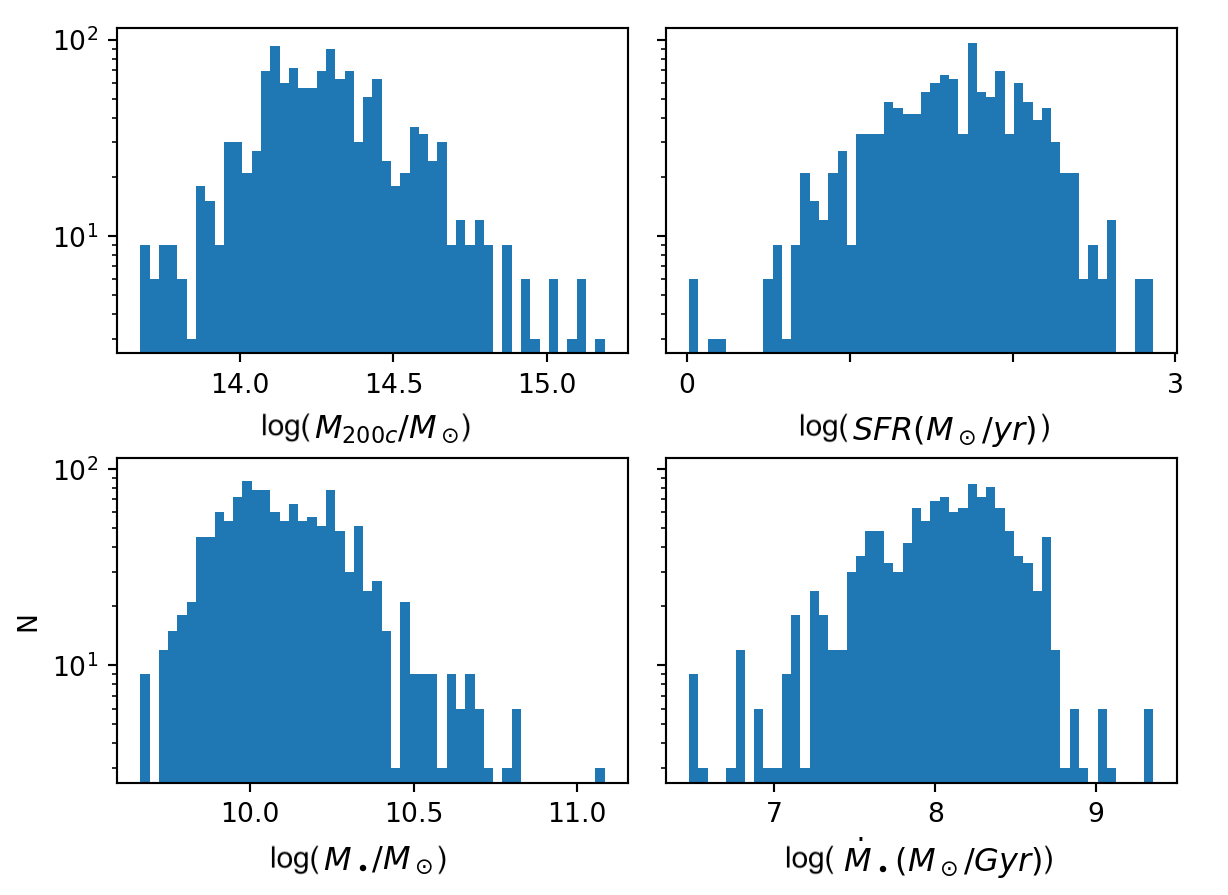}
    \caption{Properties of the galaxy clusters used in the training sample. All clusters are drawn from the TNG300 simulation \citep{Pillepich2018a-TNG}. The sample has a mean (median) virial mass $M_{200c} = 1.6 (1.3)\times10^{14}M_\odot$. The star formation rates, SMBH masses (which trace the cumulative AGN feedback over the SMBH history) and the instantaneous SMBH accretion rate all follow nearly log-normal distributions. The instantaneous stellar feedback and AGN feedback rates, which are tied to the SFR and SMBHAR, therefore span three orders of magnitude each. Besides cluster merger and accretion activity, these are expected to be the major contributors to the total X-ray luminosity.}
    \label{fig:cluster_props}
\end{figure*}

The properties of the galaxy clusters are shown in Fig \ref{fig:cluster_props}. The star formation rates, supermassive black hole (SMBH) masses, and SMBH accretion rates all follow nearly log-normal distributions, representing an unbiased selection in the amount of stellar feedback and instantaneous and cumulative AGN feedback in the sample clusters. These feedback processes are expected to be the major contributors, besides gravitational potential, to the X-ray luminosity of a cluster. 

We randomise the order of the images and split them into 80\% training, 10\% validation, and 10\% testing sets. The machine learning model is thus trained on 1826 images, unsorted by any cluster property; the model is validated at each step with 228 image pairs. We remind the reader that the training never aims to reduce the validation loss, only the training loss. Computing the validation loss on the fly allows us to assess whether the model is overfitting, i.e. learning features that only pertain to the training sample. All the results presented in this paper are for the test sample, which the network has never seen in the training process. 

Finally, the neural network is trained to reproduce the gas maps given the dark matter maps as input. Cost-minimising algorithms perform best when the inputs and outputs span a somewhat uniform range between 0 and 1. We test two different normalizations. The first, which we call `minmax, maps all the pixel values to the space between (0,1). The second, which we call `4$\sigma$', ensures that the (0,1) range contains $\mu \pm 4-\sigma$, where $\mu$ and $\sigma$ are the mean and standard deviation of the pixel values, respectively. This removes about 0.2$\%$ of the pixels but allows the remaining values to fill a much greater portion of the training parameter space.  

When applying the model trained on FP maps to maps from the DMO simulation, it is crucial to remember that the dark matter halos have slightly different structure in the absence of baryons. In other words, the minima, maxima, mean and standard deviation of the dark matter mass maps are different between the two runs. The model, however, assumes a transformation from the (0,1) to physical space based on the FP maps. Therefore, we use the parameters from the FP maps to normalise the dark matter maps from the DMO simulation as well, before passing them through the trained model. 

\begin{figure*}
\begin{center}
\includegraphics[width=0.6\textwidth]{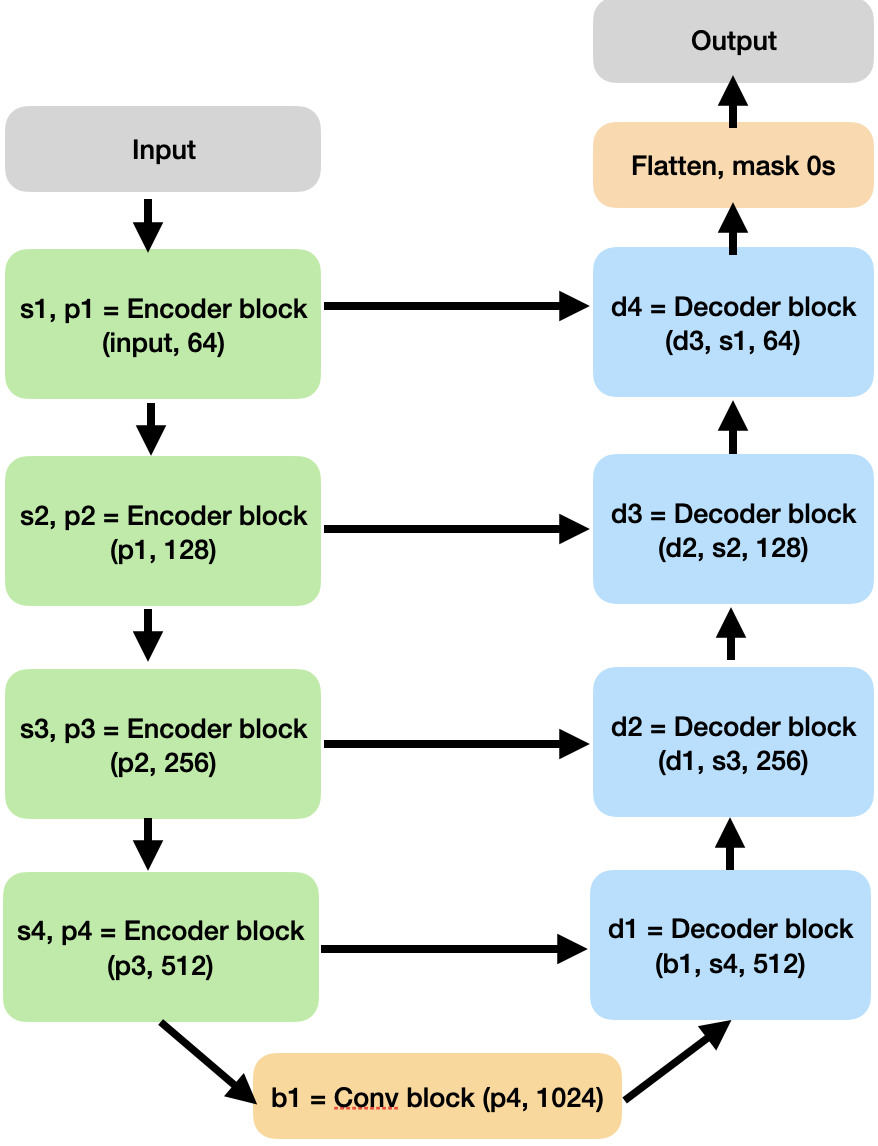}
\end{center}
\caption{Architecture of the U-Net algorithm. This is an image-to-image deep learning architecture, which has proved very successful in the fields of image colorisation and super-resolution painting. The contracting path is shown on the left, and the expanding path on the right; horizontal arrows indicate skip connections. Each convolution block includes a 2D convolution with a kernel size of (3,3), a batch normalization, and a ReLU activation, with the whole sequence repeated twice. The encoder block consists of a convolution followed by a MaxPool of kernel (2,2). Each layer extracts features on ever larger spatial scales, reducing the image to a sparser representation. The decoder is an inverse 2D convolution, a concatenation with the corresponding contracting path layer, and another convolution. At each step, therefore, it expands the sparse representation of the previous layer, while concatenating with the corresponding layer of the contracting path ensures that features are reproduced not just statistically, but at the same locations. }
\label{fig:arch}
\end{figure*}

\subsection{The neural network}
\label{sec:ml}
The task of baryon painting can be framed as an image-to-image task, where the input is the projected $N$-body simulation results and the output is a 2D observable (e.g., X-ray surface brightness map) or other desired 2D output (e.g., projected gas mass, stellar mass, star formation rate) that is derived from a full-physics simulation.  The U-net architecture \citep[e.g.][]{Ronneberger2015} is a class of image-to-image deep learning algorithms that is very popular in the field of medical imaging, due to its ability to capture features on a variety of scales.  U-nets have been used to automate the process of image segmentation \citep[e.g.,][]{Ronneberger2015}, super-resolution image reconstruction \citep[e.g.,][]{Mao2016, Yang2016}, and image colorization \citep[e.g.,][]{Zhang2016}. Painting baryons is analogous to the task of image colorization; the input gray-scale image is the dark matter map, and the colorised counterpart is the baryonic image, which can have multiple colors, often refererred to as ``channels'' in the ML literature.  U-nets have already been applied to a variety of image processing tasks in astronomy \citep[e.g.][]{Giusarma2019, Jeffrey2020, Vojtekova2021}.

A U-net is a subclass of deep convolutional neural networks, which typically employ a series of convolutional and pooling layers to extract features from the input image.  U-nets reduce the input image to a sparse representation, and then expand the sparse representation to create an output image.  To retain information about small-scale features, U-nets use skip connections that append blocks of similar sizes across the model.  
Figure \ref{fig:arch} shows a schematic of the U-net used in our baryon painting model. The contracting path comprises a series of convolutional and pooling layers, shown as green encoder blocks that reduce the input image to a sparse representation. The expanding path comprises a series of deconvolutional layers, shown as blue decoder blocks that reconstruct the output image. Horizontal black arrows show the skip connections, the feature that distinguishes U-Nets from a more traditional encoder architecture. These skip connections append each layer in the expanding path with the similarly sized outputs from encoder blocks in the contracting path.  These connections ensure that spatial fidelity is not lost in the convolution process and that the output preserves the small-scale spatial structure of the input.  Full architecture details are given in Appendix \ref{sec:fullarchitecture}.

 \begin{table}
\centering
    \begin{tabular}{l|c|c}
        \textbf{Name} & \textbf{Normalization} & 
        \textbf{Mask DM=0 pixels}\\
        \hline
        minmax & minmax & No \\
        minmax-mask & minmax & Yes \\
        4$\sigma$-mask & 4$\sigma$ & Yes \\
    \end{tabular}
    \caption{For each of the baryonic properties in Table \ref{tab:models}, we trained the algorithm three times with different choices of input normalisation and output masking, as named here.}
    \label{tab:models2}
\end{table}

Our U-net model is modeled on an example from the Keras team \footnote{\url{https://github.com/keras-team/keras-io/blob/master/examples/vision/oxford_pets_image_segmentation.py}} and is described in detail in Table \ref{tab:archfull} implemented in Keras \citep{chollet2015} with a Tensorflow \citep{45381} backend. We use a ReLU activation \citep{Agarap2018} for all but the final output, which instead uses a tanh activation.  We utilize ``same'' padding for each convolutional layer.  The model has $\sim$31 million free parameters and is compiled with an Adam optimiser \citep{2014arXiv1412.6980K} with the default learning rate and trained to minimize a pixel-by-pixel sum of the mean squared error.  The validation set is used to assess for overfitting.  Several model variations are explored; these are described in the next Section.

         


\subsection{Model variations}
For each baryonic property, we train three models. In the base model, we used the `minmax' normalization on both the input and output, and passed it through the U-net. 

\begin{table*}
    \begin{tabular}{l|l|l|l|l}
       \textbf{Name}  & \textbf{Output} & Best model & Median MPE$^\S$ ($\%$) & Mean MPE ($\%$)\\
       \hline
         $\Sigma_{DM} \rightarrow \rho_g$ & log(Projected hot gas density) & 4$\sigma$-mask & -11.4 & -16.1\\
         $\Sigma_{DM} \rightarrow T_X$ & log(Mazzotta-weighted temperature of hot gas) & minmax-mask$^*$  & -0.79 & -0.90\\
         $\Sigma_{DM} \rightarrow L_X$ & log(Projected X-ray luminosity) & minmax-mask & -1.67 & 0.88 \\
    \end{tabular}
    \caption{Summary of the best-fit models. In each case, the input is the projected dark matter mass in each pixel, $\Sigma_{DM}$. \\
    $^\S$ MPE = Mean Percentage Error, computed over the entire image. \\
    * Marginally worse MSE with 4$\sigma$-mask.}
    \label{tab:models}
\end{table*}

In the next iteration, we added a step to the end of the U-net that set all the output pixels to zero if the corresponding dark matter pixel is empty. This set of models is labelled `minmax-mask'. It encodes the physical intuition that baryons follow the dark matter, and there should not be emission or gas mass in the absence of dark matter. U-net algorithms, and convolutional neural networks in general, operate by identifying features on different scales, and therefore may not perfectly capture boundaries, especially when they vary dramatically between training images. If a lot of empty pixels are erroneously assigned even small, non-zero values, and the algorithm is minimising the mean squared error, it compensates by underestimating many of the remaining pixels.  

Lastly, we changed the normalization to $4\sigma$ (see \S \ref{sec:sims}), and retained the mask. The `minmax' normalisation can result in the training values filling a very small region of the (0,1) space, in order that a very small number of outlier pixels are included in this range. By definition, $\mu \pm 4-\sigma$ excludes only 1 in 15,787 pixels, or 0.006$\%$, but as the right column of Fig \ref{fig:pixels} shows, this allows the remaining values to fill the (0,1) space much more evenly. This allows the training to capture smaller differences between the true and predicted values. On the other hand, the densest, hottest pixels may contribute disproportionately to the total emissivity of the ICM, and it is possible that excluding them may cause significant biases in the prediction.

For most models, we took the logarithm of both the input and output quantities before the renormalisation step. Since the temperature maps have limited dynamical range, we also trained a model to reproduce the temperature in linear, rather than log, space; we call this model $T_{lin}$.

\section{Results} 
\label{sec:results}

Here, we present the results of our trained models and highlight technical lessons. We then apply the trained model to simulations of progressively lower resolution, to quantify the systematic effects of resolution and baryons on the cluster X-ray luminosity function and scaling relations. 

\subsection{Predicting baryon maps from dark matter mass in the FP simulations}

First, we trained the model to predict baryonic maps - projected gas mass, spectral-weighted projected temperature, and X-ray surface brightness - given the projected dark matter mass as input. While the gas density is not a direct observable, it is a key determinant of the X-ray luminosity; once the temperature is measured, the gas density profile can be reconstructed from surface brightness maps. For each input-output pair in Table \ref{tab:models}, we trained three models with the parameters in Table \ref{tab:models2}. 

\begin{figure*}
    \centering
    \includegraphics[width=\textwidth]{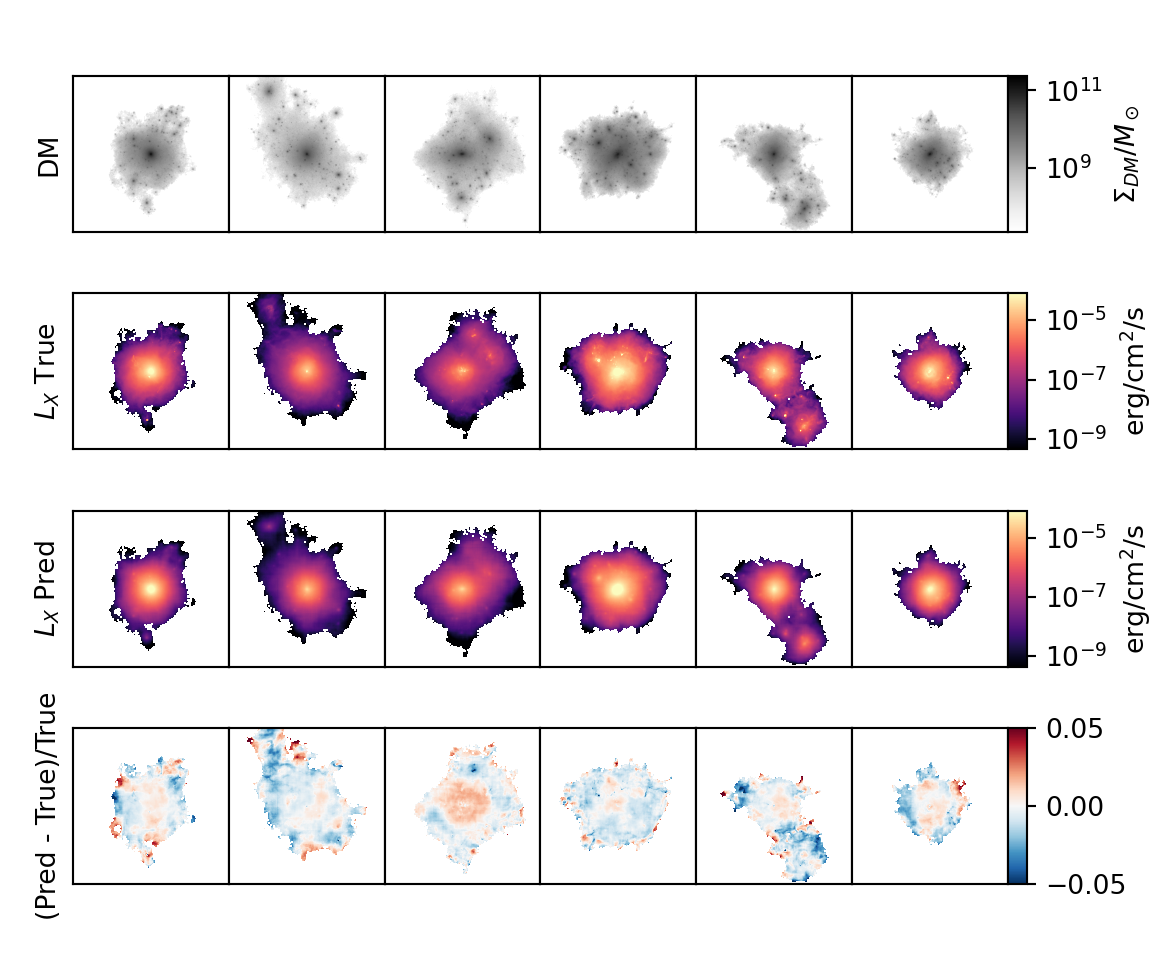}
    \caption{Results of a model trained to reproduce the projected X-ray flux given the dark matter maps as input. The projected dark matter mass density is shown in the top row, followed by true (second row) and predicted (third row) X-ray flux. The bottom row shows the fractional error between the true and predicted gas maps in the unphysical training space. Converted back to the physical space, the errors on a pixel-to-pixel level can be factors of several, but when summed over the entire image, the median (mean) MSE is 6.11 (13.36)$\%$.}
    \label{fig:mass_lx_2d}
\end{figure*}

The best-fit model for the X-ray luminosity is shown in Fig \ref{fig:mass_lx_2d}. The top row shows the dark matter map, the second shows the true X-ray luminosity, the third row shows the model prediction, and the bottom row shows the fractional difference between the true and predicted gas properties in the unphysical, normalised training space where the loss was minimised. Immediately, there is excellent agreement in the magnitude as well as structure between the true and predicted maps. While the models were trained to minimise the MSE, Table \ref{tab:models} shows the mean percentage error MPE = ($L_{X, \rm pred} - L_{X, \rm true})/L_{X, true}$. Here, $L_X$ is the sum of surface brightness over each image, reconstructed to physical units, and its median value for the best-fit model is just -1.67$\%$; the temperature performs similarly well. The gas density is underpredicted at the $10\%$ level even in the best fit model.

The first column shows one of the more dramatic - although still minor - cases of model underprediction, which is mostly in infalling groups at the outskirts. The X-ray luminosity maps were generated using only the hot gas, i.e. $T > 3\times10^5K$ and $\rho_g < 5\times10^{-25}{\rm g/cm^3}$, criteria which are unlikely to be met by most of the gas in low-mass groups. Groups that are infalling into a cluster environment can be significantly shock heated to uncharacteristic degrees in the X-ray. Such early-stage group infalls are rare in our sample of 761 clusters, and we presume that this kind of system would be predicted better by an algorithm trained on a larger, more diverse cluster sample. 

The gas density and temperature maps (Figs. \ref{fig:mass_rho_2d} and \ref{fig:mass_temp_2d})) perform remarkably well, although they predict less substructure than actually produced in the simulations. This is likely the highly non-linear effect of radiative cooling, which does not follow directly from the instantaneous gravitational potential. Nevertheless, the median reconstruction errors for these two models are 11.4$\%$ and 1.02$\%$, respectively. 
   
\begin{figure*}
    \includegraphics[width=0.8\textwidth]{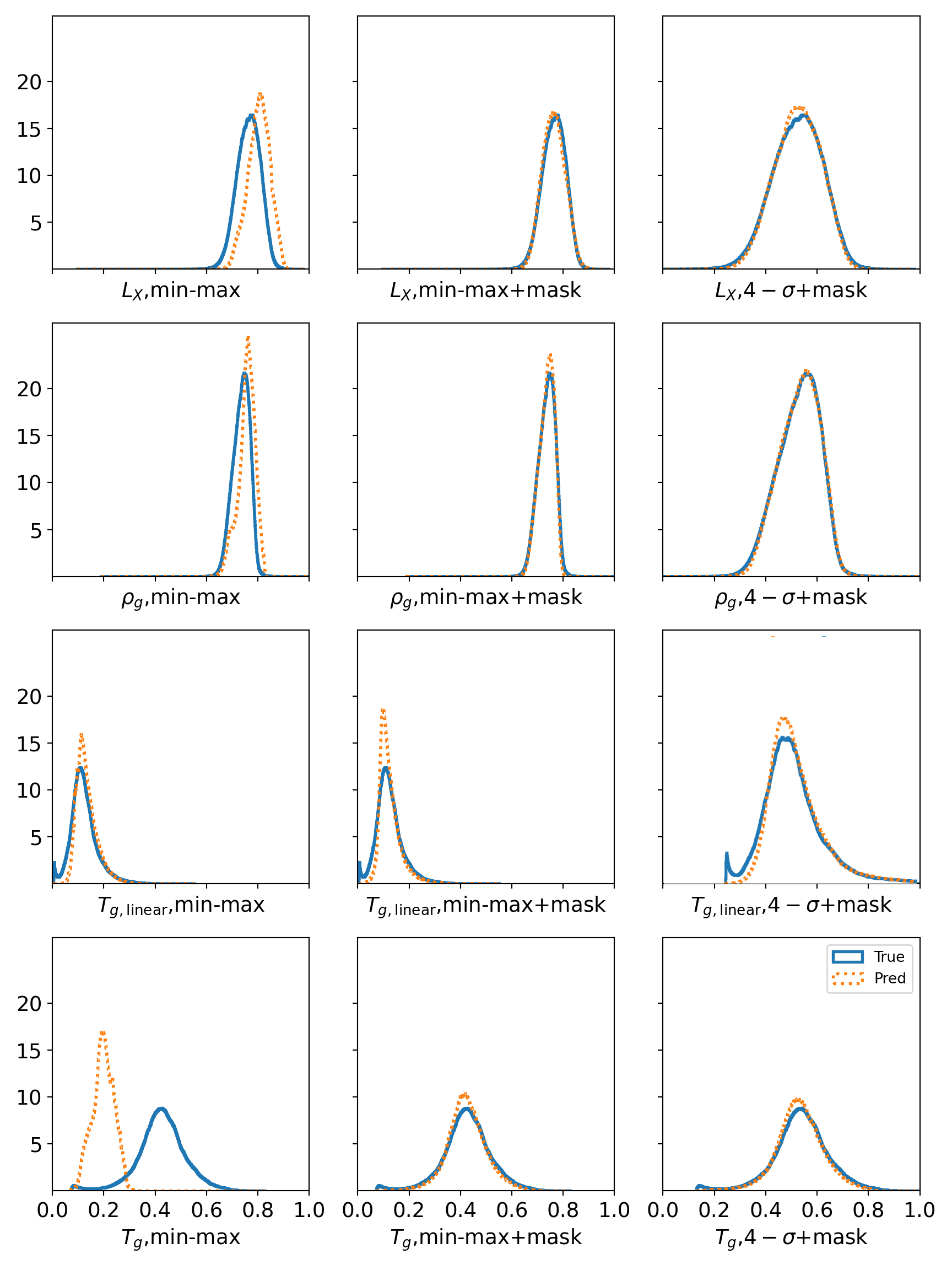}
    \caption{PDF of pixel values of the true (blue) and predicted (orange) images for the various models, in non-physical units between 0 and 1. In the left panels, the inputs are transformed so that the logarithm of their values fits in the (0, 1) space; this is the baseline or `minmax' model. Masking out the empty regions of the dark matter maps, as shown in the middle row, improves the similarity between the true and predicted distributions; this is the `minmax-mask' model. In the right panels, in addition to masking the outputs where input is 0, the input normalization is changed so that the (0,1) range contains ($\mu\pm4-\sigma$) of the values rather than the full range of input values, which otherwise may fill a very limited range of the training space. This is the `4$\sigma$-mask' model. All the models tend to underpredict the values of the brightest pixels, and conversely overpredict the faintest pixels; this is the well known problem of bias towards the mean in regression models.}
    \label{fig:pixels}
\end{figure*}

\begin{figure*}
    \includegraphics[width=\textwidth]{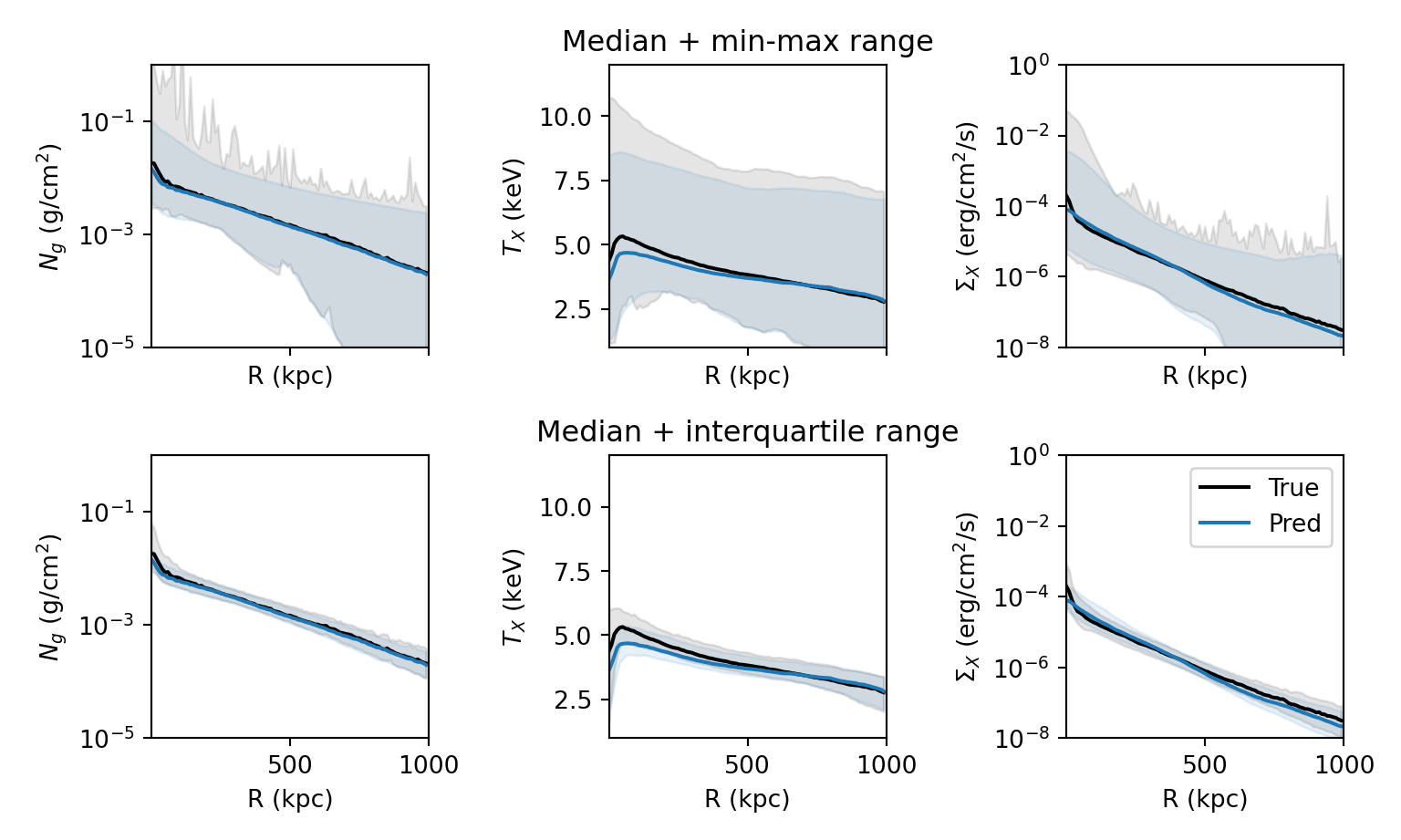}
    \caption{Median profiles of the azimuthally averaged quantities predicted from the best-fit models for each property. Profiles were predicted for only the test sample. In the top row, the shaded region represents the full range of the true and predicted profiles; in the bottom row, it shows the interquartile range, i.e. between the 25th and 75th percentiles at each radius. This shows that while some of the true profiles show some spiky features at all radii, these are outliers that do not appear with the interquartile range.}
    \label{fig:profiles}
\end{figure*}

\subsection{Effect of input normalisation and masking}
Fig \ref{fig:pixels} compares the distributions of the pixel values in the true and predicted images for each of the twelve trained models. The left and center columns use the `minmax' normalisation, while the right uses `4$\sigma$'; the center and right columns additionally mask the outputs in all the pixels where the input (DM mass) is 0. 

The base `minmax' model, in the left column, performs best when predicting $T_{lin}$, recovering the mean and only underestimating the standard deviation of the pixel values by $\sim25\%$. For all the other outputs, however, the predicted images have a different mean from the ground truth - the density and X-ray emissivity are systematically overpredicted, while the temperature is underpredicted. This reflects the tendency of CNNs to bias predictions towards the mean. A very small number of very dense pixels contain very high gas density and X-ray emissivity, and due to radiative cooling some of these pixels will have very low temperatures. Since they are so unlikely, however, the model will assign them values that are closer to the mean; to compensate, it will also adjust the predictions in the rest of the distribution so that the MSE over the entire image is low. For the gas density and X-ray emissivity, this means that densest/brightest pixels are underpredicted, but the remaining pixels are denser/brighter on average; similarly, the coldest pixels are predicted to be hotter than the true value, but to compensate, many of the remaining pixels are predicted to be colder than they ought to be. In all cases, the effect is dominated by the highest density pixels. 

Masking out the output pixels where the input pixels are 0, as shown in the middle panel, limits the number of pixels where the CNN can compensate for this bias. Where earlier it could assign a small, non-zero value to the 0 pixels, which would add up to counter the predicted deficit in the densest pixels, it now cannot. This forces it to improve the prediction in the space where there actually is dark and baryonic matter. Hence the middle column of Fig \ref{fig:pixels} performs significantly better than the left column for all output properties. 

Some properties, like the gas density and X-ray flux, have a small number of outliers that have uncharacteristically high or low values. If all the pixels are required to live in the (0,1) space, most of the pixels available for training actually live in a small fraction of that space. The $4-\sigma$ normalization, by dropping just $0.06\%$ of the pixels, allows the remaining pixels to fill the (0,1) space much more evenly, as shown in the right column of Fig \ref{fig:pixels}. This should, in principle, allow a smoother mapping between the input and output quantities, since the dynamic range of each is effectively expanded. The trade-off is that a few very faint or very bright pixels fall outside the training domain. The gas density shows a clear preference for the $4\sigma$ normalisation, whereas for the others, it is not clear by eye whether the peakier predicted distribution with the `minmax' normalisation offsets the other differences between the distributions. When integrated over the entire image, however, Table \ref{tab:models} shows that the temperature and X-ray flux are better predicted with the `minmax' normalisation with output masking. 

\begin{figure*}
    \includegraphics[width=\textwidth]{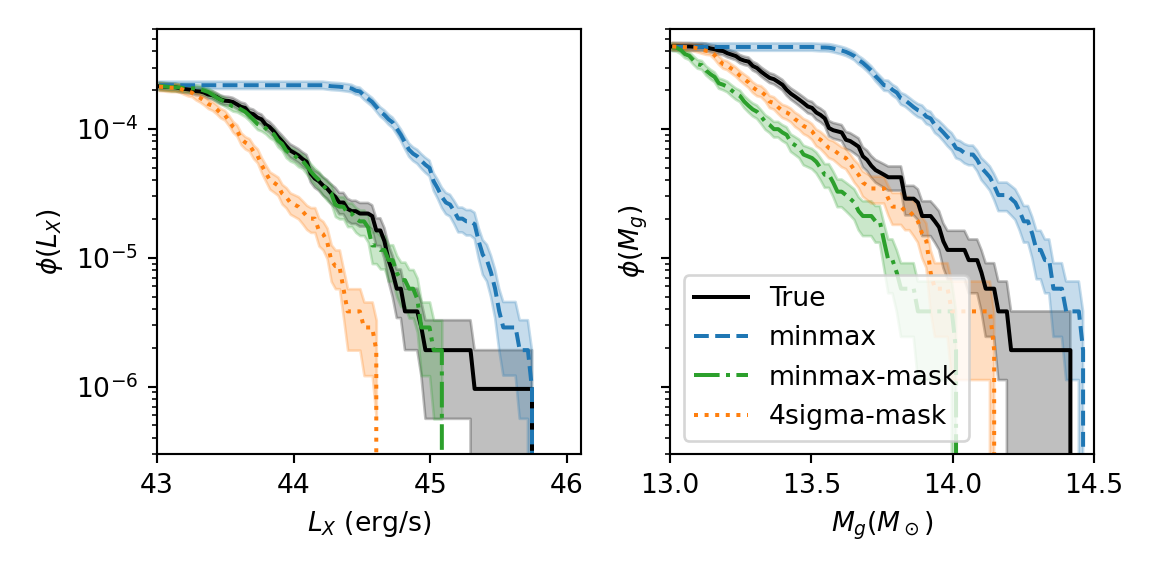}
    \caption{The true (black, solid) and predicted distribution function for the cluster X-ray luminosity (left) and gas mass (right), computed as a simple sum over the images. The shaded bands indicate Poisson uncertainties. The gas mass function is best reproduced using the $\mu\pm4-\sigma$ mask on the inputs (orange, dotted), whereas $L_X$ performs better if the renormalised dark matter projected mass is allowed to fill the entire (0,1) space (green, dot-dashed); this is likely because the few dense pixels contribute super-linearly to the luminosity but only linearly to gas mass. Masking out the empty pixels improves all the models.}
    \label{fig:functions}
\end{figure*}

\subsection{Radial profiles}

Fig \ref{fig:profiles} reduces the 2D images to azimuthally averaged radial profiles, and shows the excellent agreement between the true (black) and predicted (blue) values. The solid line in each plot shows the median profile at that radius out of the test sample of 228 clusters. In the top rows, the shaded region ranges from the minimum to the maximum value in each radial bin; in the bottom row, the shaded region is the interquartile range, i.e. from the 25th to the 75th percentile. The comparison shows that while there are some spiky features in some of the true profiles, these are rare outliers that do not appear in the interquartile range. Image-to-image baryon painting thus also reproduces the radial profiles of the intracluster medium and most of its diversity. 

\subsection{Predicting X-ray and gas mass distribution functions}
Summing up the X-ray flux and gas column density over each of the images yields a single value for X-ray luminosity $L_X$ and and gas mass $M_g$ for each cluster. We compare the distribution functions of these  properties to the true values from the FP simulation in Fig \ref{fig:functions}. This is not something the model was explicitly train to reproduce - it only learned the mapping between a given dark matter map and its baryonic counterpart. The dark matter mass function does constrain the distribution functions to first order, but because galaxy clusters exhibit a large scatter in ICM properties at fixed halo mass, this does not have to translate into a correct distribution function of baryonic properties. We further emphasise that while these are integrated quantities, they do in fact make use of the 2D information. If instead we had trained a model where the dark matter mass was provided as a single value per halo, the output could not have the scatter seen in the “true” simulated population. Using 2D images is equivalent to constructing an input vector that contains the halo mass, mass accretion history, shape parameters, redshift, and other quantities that drive the scatter in cluster observables. 

The best fitting models \textemdash{} `minmax-mask' for $L_X$ and `4$\sigma$-mask' for $M_g$ \textemdash{} produce cluster distribution functions that align very well with the true values within the Poisson 1-$\sigma$ uncertainties. Without masking the DM-free pixels, both distribution functions are biased high; excluding the rare, but very bright, pixels with the $4-\sigma$ normalisation has a greater effect on $L_X \propto n_g^2$ than $M_g \propto n_g$. Despite training over less than one decade in halo mass, we reproduce the cluster luminosity function over three orders of magnitude. 

\subsection{Painting a DMO simulation with a model trained on full-physics simulations}

As shown in Fig \ref{fig:dm_vd_fp}, the dark matter halos produced in the DMO runs look different from their FP counterparts. This is partly due to the stochasticity introduced by the $N$-body problem; furthermore, baryons introduce a variety of systematic but highly non-linear effects on their dark matter halos in ways that depend on the halo mass and the baryonic feedback prescriptions \citep{Kochanek2001,Pedrosa2009,Duffy2010}. We therefore expect that training our model on DM properties extracted from the FP simulation will be biased compared to if we had DM halos from a simulation without baryons, if somehow the stochasticity of the $N$-body problem could be removed.

\begin{figure*}
    \centering
    \includegraphics[width=\textwidth]{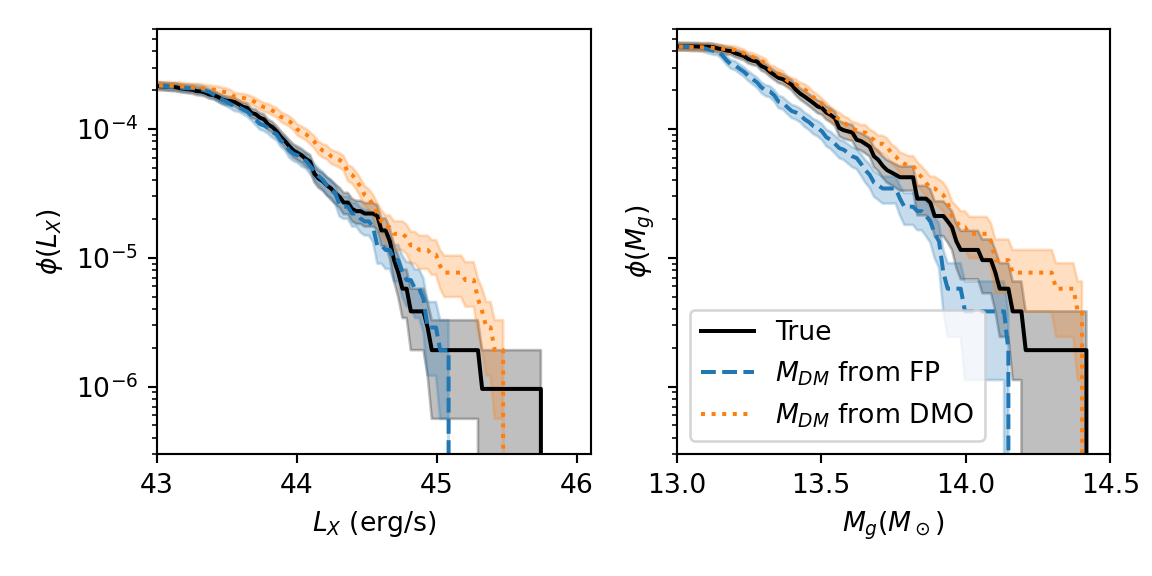}
    \caption{The X-ray luminosity (left) and gas mass (right) distribution functions predicted using the corresponding best-fit models and dark matter maps from the FP (blue) and DMO (orange) simulations. The shaded regions, again, indicate Poisson noise. There is excellent agreement between the predicted and true distribution functions, which the model was not trained to reproduce. Note further the three orders of magnitude in $L_X$ predicted from one order of magnitude in $M_{DM}$.}
    \label{fig:functions-dmo}
\end{figure*}

The next step, therefore, is to check whether models trained on dark matter maps from the FP simulation perform adequately when applied to dark matter maps from the DMO simulation. The model assumes that the mapping between the physical and training (0,1) space is based on the FP simulation; therefore, we use the same numbers to renormalise the DMO maps to the training space. Since the input is now from an $N$-body simulation, there is no ground truth for the baryonic properties. Nevertheless, we can test whether the model can predict the correct distribution function of baryonic observables. 

Fig \ref{fig:functions-dmo} shows that the models do indeed succeed in reproducing the FP distribution functions even when using inputs from the DMO run. The X-ray luminosity function is slightly overestimated at $10^{43.5-45}$ erg/s, whereas the gas mass distribution from the best-fit (`4$\sigma$-mask') model agrees remarkably well with the FP distribution within the entire training domain. Both models slightly overpredict the distribution function when applied to the $N$-body simulation, and because the $\Phi(M_g)$ was slightly underpredicted using inputs from the FP run, the two effects cancel out, producing a very good agreement between the predicted gas mass function from DMO and the true distribution from FP. This captures a well known difference between $N$-body simulations and their full-physics counterparts. The former systematically produce more ultra-dense pixels, which in the FP runs get smoothed out by baryonic feedback. We checked that using the `$4\sigma$-mask' overcompensates for this effect in $\Phi(L_X)$, and slightly underpredicts the galaxy cluster X-ray luminosity function. 

In practice, both these biases can be computed and accounted for. For each bin in $L_X$ or $M_g$, we can compute the bias between the true and predicted distribution functions. This correction factor is the net effect of the CNNs bias towards the mean, and the tendency of $N$-body simulations to produce more ultra-dense clumps than their full-physics counterparts. 

\begin{figure*}
    \includegraphics[width=0.95\textwidth]{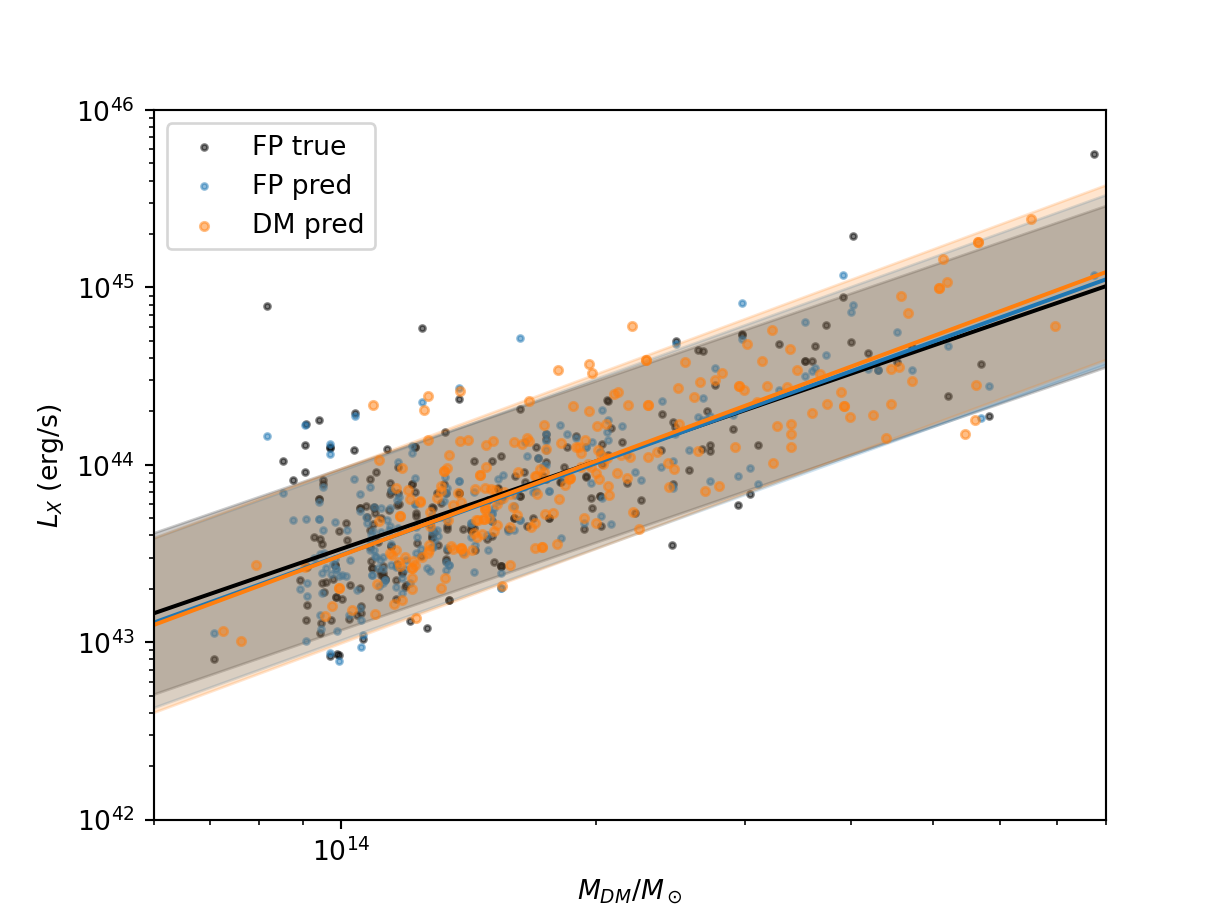}
    \caption{The scaling relation of the total dark matter mass $M_{DM}$ vs the cluster luminosity $L_X$. Black points show the true values from the FP simulation, blue are the predictions for dark matter maps from the test set of the FP simulation, and orange points are predictions for the DMO simulation. The corresponding lines show the best-fit scaling relations. While the predicted slopes are slightly steeper, they lie well within the 1-$\sigma$ uncertainties of the true model. }
    \label{fig:scaling}
\end{figure*}

\begin{figure*}
    \includegraphics[width=\textwidth]{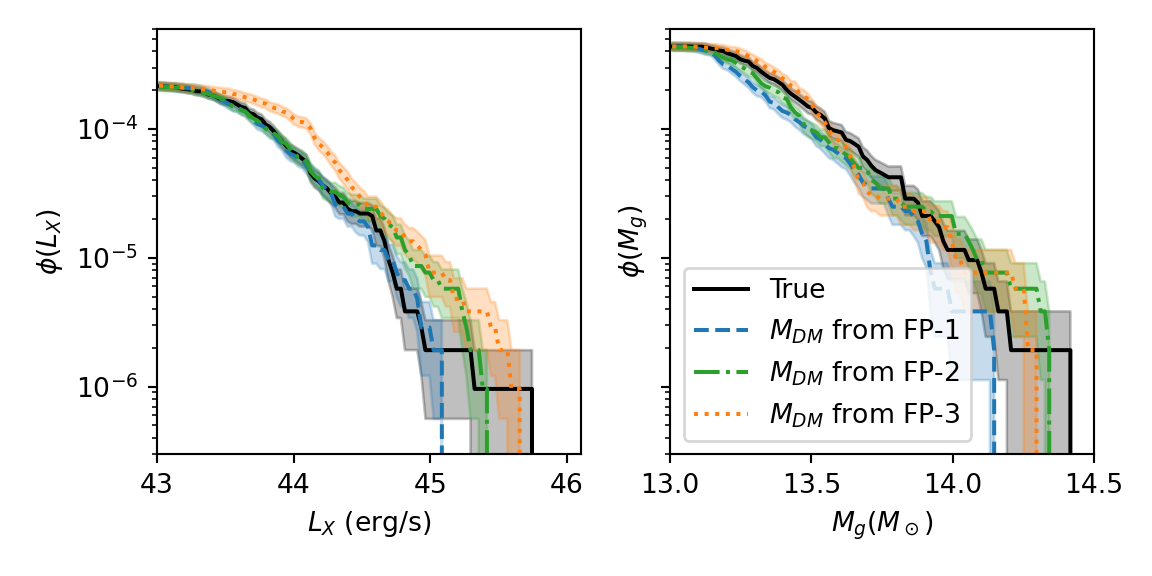}
    \caption{Distribution functions of the X-ray luminosity and gas mass predicted for dark matter mass maps from the medium (FP-2) and low (FP-3) resolution runs of TNG300, which have mass resolutions 8 and 64 times lower than FP-1, on which the training was performed. Predictions for FP-2 agree with the ground truth for luminosities below $10^{45}$ erg/s and gas mass below $10^{14}M_\odot$; for more massive/brighter clusters, the model trained on FP-1 over- (under-) predicts the $L_X$ ($M_g$) distribution function.}
    \label{fig:restest}
\end{figure*}

\subsection{X-ray - mass scaling relation}
Fig \ref{fig:scaling} shows the scaling relation between the dark matter mass and X-ray luminosity in the FP simulation, compared with predictions for the FP test sample and for the DMO sample. The best-fit scaling relations are remarkably similar. Further, the U-net naturally predicts the scatter in these scaling relations, which are captured in the detailed spatial structure of the cluster and therefore the intracluster medium. While the scatter may physically source from gas clumping, star formation, and stellar and AGN feedback, the success of this model tells us that all these physical processes are correlated with the detailed spatial structure of the halo; baryons follow the dark matter, and denser regions correspond to more active baryonic feedback. The U-net architecture is therefore capable of numerically describing these spatial correlations and reproducing the observed richness of cluster properties. 

The fact that the machine learning model can reproduce the X-ray luminosity function, $L_X-M_{DM}$ scaling relation and the scatter therein, means that it can be applied to any DMO simulation to paint the Illustris-TNG physics model onto it. Predicting a single property for $\sim2700$ DMO images on 2 GPUs took 73 seconds. This means that every DMO simulation today can be converted into a large-volume full-physics realisation of Illustris-TNG. Similar models can, and must, be trained on other simulations like EAGLE \citep{Schaye2015} and SIMBA \citep{Dave2019} to numerically capture the existing uncertainty between theoretical models of galaxy evolution.  

\subsection{Resolution effects}
Lastly, TNG300-1 has an exquisite sub-kpc resolution that is much finer than most large-volume $N$-body simulations. To apply this model to existing $N$-body simulations, it is important to assess how it performs on dark matter maps from lower-resolution simulations. Fig \ref{fig:restest} shows close agreement between the predictions for FP-2, which has 2 times lower resolution in space (and therefore 8 times lower in mass) than the training sample, and the ground truth from FP-1, for all but the brightest and most massive galaxies. 

Degrading the spatial (mass) resolution of the dark matter simulation by a further factor of 2 (8) produces an X-ray luminosity function biased high; this does not seem to hold true for the gas mass function. This is consistent with known effects of resolution on dark matter density maps - low resolution simulations will wash out small-scale dense peaks, which are rare but contribute significantly to both luminosity and gas mass. Since the best-fit X-ray luminosity model used the `minmax' normalisation, it is more affected by these very rare, very dense pixels; the `4$\sigma$' normalisation simply excludes them from the training domain. 

We conclude that our model has some potential for super-resolution baryon painting, but we advice caution when generating X-ray luminosity functions from low-resolution $N$-body simulations. 

\section{Discussion and Future work}
\label{sec:disc}
\subsection{The impact of the TNG physics model}
As noted above, our study enables painting the TNG physics model onto $N$-body simulations using any choice of cosmological parameters. It is likely, however, that different choices of feedback from AGN and stars could produce different ICMs \citep[e.g.,][]{Delgado2023}. It will be crucial to repeat this experiment with other simulations including a significant number of galaxy clusters, such as C-EAGLE and \textit{Magneticum}. The spread between the different simulations are representative of the current theoretical uncertainty in the effect of baryonic processes on dark matter halos. 

\subsection{Is the model learning cosmology?}
One concern that we cannot rule out at the moment is that the mapping that the models have learned actually describe the \citet{Planck2016} cosmology implemented in the TNG simulations, and that the mapping looks fundamentally different in other cosmologies. We do not expect this to be the case, since the baryonic physics depends only on local properties and their history over time, and this is captured in the instantaneous dark matter maps. However, this can be robustly tested with suites of simulations like \textit{Magneticum} \citep{Dolag2016} which, despite lower resolution, ran several simulations with identical physical models and different cosmologies. Given the complexity of training on a suite of simulations that is not currently in the public domain, we leave this test to future work. 

\subsection{Multi-wavelength predictions}
While the ongoing \textit{eROSITA} all-sky survey motivated an emphasis on the X-ray luminosity function, the coming years will see game-changing SZ surveys such as CMB-S4 \citep{Abazajian2019} and the Simons Observatory \citep{Ade2019}; at radio wavelengths, LOFAR \citep{Cassano2010, Savini2019}, MeerKAT \citep{Knowles2022} and ASCAP are already setting an exciting stage for SKA \citep{Gitti2018}. Modeling observations at both sub-mm and radio wavelengths must be done very carefully to account for systematics in the interferometric observation pipeline. In addition, radio synchrotron emission depends not only on the bulk properties of the gas, but also relic populations of relativistic electrons and the details of cosmic ray feedback. This makes radio observations harder to link to cosmology than X-ray and SZ. However, calculating $Y_{SZ}$ from the simulations is simple, and pipelines do exist for mocking its observations with real telescopes. Applying our technique to the $Y_{SZ}$ function of galaxy clusters could allow us to catch unnoticed systematics in X-ray surveys, or vice versa. 

\subsection{The effect of adding dynamical, redshift and SMBH information}
We trained additional models to use additional input channels besides the projected dark matter mass. First, since the intracluster medium is known to be sensitive to the mass accretion history of the cluster, we added maps of the average velocity magnitude of the dark matter particles. Second, since AGN feedback is known to regulate the thermodynamics of the ICM, we created an additional channel that included the positions and instantaneous accretion rates of all the SMBH within the cluster.

Somewhat surprisingly, neither of these models performed better than the base model. This is likely because dynamical and heating effects are local, and lost in projected images. Further, since AGN reside preferentially in the densest regions of clusters, information about their effect may have been encoded in the central dark matter distribution of each cluster. We remind the reader again that direct emission from the AGN is not included in our maps; this would be a helpful next step, which would allow the direct interpretation of X-ray luminosity functions from lower-resolution surveys like \textit{eROSITA} where deconvolving the broad PSF of the AGN can prove challenging. 

Finally, we trained models using only clusters at fixed redshift, i.e. one model each for z = 1.0, 0.5, 0.3 and 0. If there is additional information encoded in the cluster redshift, these models ought to outperform the one where clusters at all redshifts are treated equally. We find that this is not the case. If there is additional information in the cluster redshift, it is not sufficient to compensate for the reduced training size. This could nevertheless be a useful feature to incorporate into future models as and when larger training sets become available.

\subsection{Improving super-resolution baryon painting}
Fig \ref{fig:restest} showed that if the model is applied to dark matter maps from simulations that have far lower native resolution than the training set, the predicted distribution functions have a different shape from the ground truth. This is because both over- and under-densities on the smallest scales are smoothed out, such that they are no longer captured by the convolutional filters trained at higher resolution. Super-resolution models have successfully been trained using CNN architectures \citep{McCann2017, Fukami2019, Soltis2022}, and are a logical next step in this line of study.

A particularly promising approach to super-resolution baryon painting is using known physical relations between the dark matter and baryonic processes, which are described in the simulation sub-grid models as differential equations (DEs). Solving the equation is equivalent to minimising the difference between the two sides of the equation. Adding this difference to the loss function of a machine learning model lies at the crux of Physics-Informed Neural Networks \citep[PINNs, e.g.,][]{Raissi2019, Cai2021}. 

\section{Conclusions}
We trained a U-Net convolution neural network (CNN) to paint X-ray surface brightness, projected gas density and spectroscopic-like temperatures onto maps of projected mass from dark-matter-only simulations. 
\begin{itemize}
    \item We find that this method works very well, with median fractional errors (on the test sample) of -11.4$\%$ for the gas column density, $-0.79\%$ for spectral-weighted projected temperature and -1.67$\%$ for X-ray luminosity. Individual pixels may vary by up to factors of a few, but the distribution over each cluster is remarkably well recovered.
    \item Using just the dark matter mass maps, the model is very successful at reproducing the baryonic structure in clusters undergoing complex mergers, where hydrostatic equilibrium cannot be assumed. It does, however, underpredict the luminosity from low-mass infalling groups, where the effect of non-gravitational heating from shocks is unusually high compared to most of the training sample. This will likely improve with larger training samples that include more mergers. 
    \item Despite being trained only to reproduce individual images, the model also reproduces the radial profiles, X-ray luminosity function and gas mass function of galaxy clusters along with almost all their inherent diversity. The model further reproduces not only the scaling relation between the dark matter mass and X-ray luminosity of galaxy clusters, but also the scatter therein. 
    \item The predictions work very well for dark matter maps drawn from the DMO simulation, even though they were trained on the FP simulation, where the dark matter structure is slightly different due to baryonic effects. The predicted distribution functions are biased slightly high, following the well-known phenomenon that the DM maps from DMO simulations contain more ultra-dense clumps than their FP counterparts, where baryonic feedback smooths them out. 
    \item The model also performs remarkably well on dark matter maps drawn from the FP-2 simulation, whose spatial (mass) resolution is 2 (8) times lower than the training sample. Further degrading the resolution results in over-predicting the X-ray luminosity function, since low-resolution simulations produce more ultra-dense pixels which in higher resolution runs would consist of several, less dense pixels. We therefore caution against using models trained at a given spatial resolution on dark matter maps from a simulation whose native spatial resolution is more than 2 times coarser.
\end{itemize} 

We conclude that U-nets are a powerful technique for learning the mapping between dark matter only (DMO) and full-physics (FP) simulations. The physical processes that alter the observable properties of galaxy clusters are correlated with the detailed structure of the dark matter, and CNNs excel at capturing such spatial correlations. The key predictive features of the dark matter maps are preserved in simulations with slightly lower resolution. This work sets the stage for further research in super-resolution baryon painting, and makes it possible to perform galaxy cluster cosmology with existing $N$-body simulations while accounting for the complex, non-linear effects of baryons on their observable properties. 

\section{Data Availability Statement}
All the data from the IllustrisTNG suite is available at \url{https://www.tng-project.org/data/}. The code for creating images from TNG300 and training the U-net, as well as all the trained models mentioned in this paper, can be found at \url{https://github.com/milchada/MLBaryonPainting}. This work uses only data in the public domain and is entirely reproducible. 

\bigskip
\begin{small}
\noindent
\textit{Acknowledgements}. We thank the anonymous referee for their very helpful suggestions. UC, JAZ, \'AB, RPK  acknowledge support from the Smithsonian Institution and the Chandra High Resolution Camera Project through NASA contract NAS8-03060. The material presented is based upon work supported by NASA under award No.~80NSSC22K0821. Helpful advice was provided by Cecilia Garraffo and the AstroAI group at the CfA, Camille Avestruz, Francisco (Paco) Villaescusa-Navarro, Antonio Ragagnin, and Joop Schaye.
\end{small}

\bibliographystyle{mnras}
\bibliography{reference.bib}

\clearpage
\appendix
\section{The full architecture of the U-Net} \label{sec:fullarchitecture}
Our U-Net was implemented in Keras with a Google Tensorflow backend. The full code can be found at \url{https://github.com/milchada/MLBaryonPainting}. The model layers are presented in Table \ref{tab:archfull}.

\begin{table*}
\centering
\begin{tabular}{l|l|l|l|l|l|l}
\textbf{Layer (type)}      & \textbf{Normalization} & \textbf{Activation}  &  \textbf{Pooling} &  \textbf{Output Shape } & \textbf{Notes}   \\               
\hline
\textbf{Contracting Path } \\
\hline
Input                       &               &       &                   &    (512, 512, 1)      &       \\    
2D convolution              & batch norm    &  ReLU &                   &    (512, 512, 64)     &       \\            
2D convolution              & batch norm    &  ReLU &  Max Pooling      &    (512, 512, 64)     &   $\mathcal{A}$    \\
2D convolution              & batch norm    &  ReLU &                   &    (256, 256, 128)    &       \\            
2D convolution              & batch norm    &  ReLU &  Max Pooling      &    (256, 256, 128)    &  $\mathcal{B}$     \\
2D convolution              & batch norm    &  ReLU &                   &    (128, 128, 256)    &       \\            
2D convolution              & batch norm    &  ReLU &  Max Pooling      &    (128, 128, 256)    &  $\mathcal{C}$     \\
2D convolution              & batch norm    &  ReLU &                   &    (64, 64, 512)      &       \\            
2D convolution              & batch norm    &  ReLU &  Max Pooling      &    (64, 64, 512)      &   $\mathcal{D}$    \\
2D convolution              & batch norm    &  ReLU &                   &    (32, 32, 1024)     &       \\            
 \hline 
 \textbf{Expanding Path} \\
 \hline
2D transpose convolution    &               &  ReLU &                   &    (64, 64, 512)     &    Output concatenated with $\mathcal{D}$   \\ 
2D convolution              & batch norm    &  ReLU &                   &    (64, 64, 512)     &       \\            
2D convolution              & batch norm    &  ReLU &                   &    (64, 64, 512)     &       \\   
2D transpose convolution    &               &  ReLU &                   &    (128, 128, 256)     &    Output concatenated with $\mathcal{C}$   \\ 
2D convolution              & batch norm    &  ReLU &                   &    (128, 128, 256)     &       \\            
2D convolution              & batch norm    &  ReLU &                   &    (128, 128, 256)     &       \\
2D transpose convolution    &               &  ReLU &                   &    (256, 256, 128)     &    Output concatenated with $\mathcal{B}$   \\ 
2D convolution              & batch norm    &  ReLU &                   &    (256, 256, 128)     &       \\            
2D convolution              & batch norm    &  ReLU &                   &    (256, 256, 128)     &       \\
2D transpose convolution    &               &  ReLU &                   &    (512, 512, 64)     &    Output concatenated with $\mathcal{A}$   \\ 
2D convolution              & batch norm    &  ReLU &                   &    (512, 512, 64)     &       \\            
2D convolution              & batch norm    &  ReLU or Sigmoid$^*$ &                   &    (512, 512, 64)     &       \\
Multiplication              &               & linear&                   &    (512, 512, 1)      &     Mask is a (512, 512, 1) \\
                            &               &       &                   &                       & input of 0's and 1's.     \\
\hline
Total params: 31,054,145 \\
Trainable params: 31,042,369\\
Non-trainable params: 11,776\\
\end{tabular}
\caption{Full architure of the employed U-Net.\\ *The activation in the final convolution layer was ReLU for the `minmax' normalisation and Sigmoid for the `4$\sigma$' normalisation, because ReLU, unlike Sigmoid, accepts inputs only in the (0,1) range.}
\label{tab:archfull}
\end{table*}

\section{Predictions from the gas density and temperature}
\begin{figure*}
    \centering
    \includegraphics[width=\textwidth]{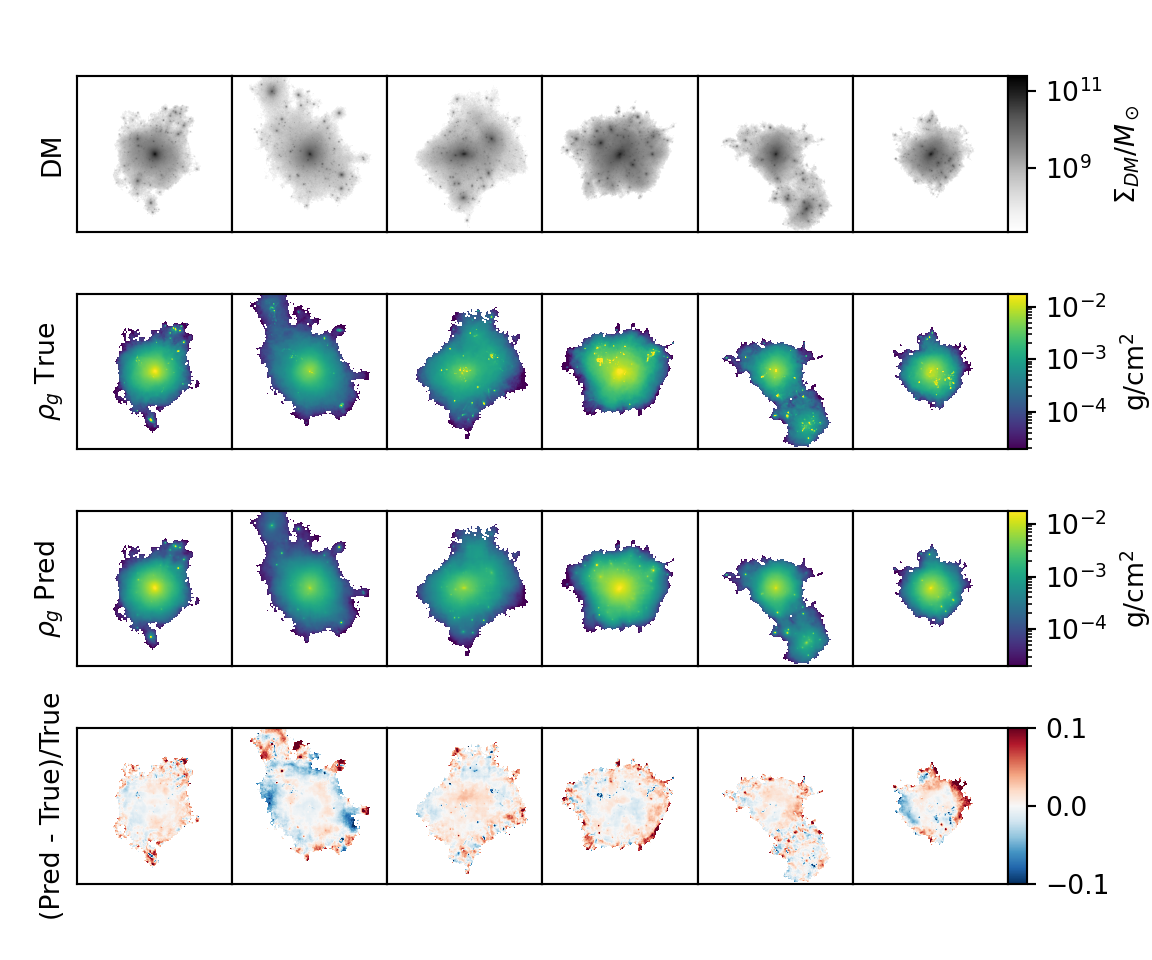}
    \caption{Results of a model trained to reproduce the projected gas density given the dark matter maps as input. The projected dark matter mass density is shown in the top row, followed by true (second row) and predicted (third row) gas column density. The bottom row shows the fractional error between the true and predicted gas maps. The MPE is at the $10\%$ level for each image.}
    \label{fig:mass_rho_2d}
\end{figure*}

\begin{figure*}
    \centering
    \includegraphics[width=\textwidth]{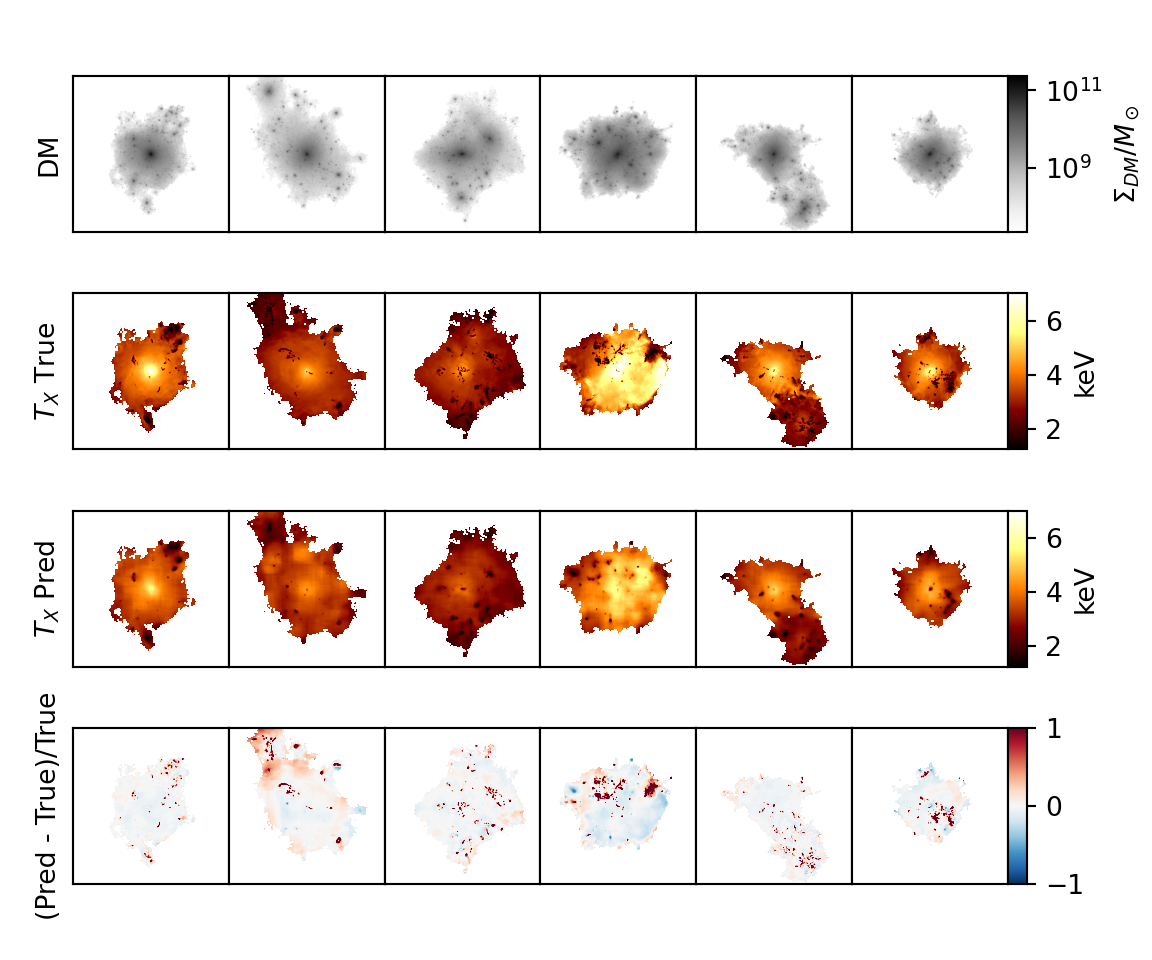}
    \caption{Same as Fig \ref{fig:mass_rho_2d}, but predicting the spectral-weighted temperature maps instead. This property has the most large-error pixels, which are systematically hotter than the ground truth. The rest of the image is systematically underpredicted to minimise MSE, so that the MPE is at the $<1\%$ level. The temperature maps show a lot of filamentary structure due to radiative cooling, which is harder to encode in the dark matter mass maps.}
    \label{fig:mass_temp_2d}
\end{figure*}

Figs \ref{fig:mass_rho_2d} and \ref{fig:mass_temp_2d} present the predictions for the best-fit models of projected gas density and spectral-weighted temperature, respectively. In the gas density map, we can see a small number of very dense pixels that were unpredicted by the model, as expected due to the bias towards the mean and slightly exacerbated by the $4\sigma$ normalisation. However, using the `minmax' normalisation for the gas density significantly increased the reconstruction error, showing that the benefit of better filling the training space outweighed the cost of excluding a few ultra-dense pixels.


\label{lastpage}
\end{document}